# Vertical Handoff Decision Based On Genetic Algorithm in 4G Network

Bijay Paikaray

**Abstract-** The rapid improvement of the mobile generations was for the purpose of supporting as many mobile devices as possible that could benefit the users at anytime and anywhere in terms of common practical applications such as internet access, video-on-demand, video conferencing system and many more applications. In this paper, a review for the mobile generations in the wireless communications is presented in order to highlight and compare the issues and challenges that are involved in each generation starting from the earlier generations along to the following generations and finally till the 4$^{th}$ Generation (4G).

The 4G wireless network is intended to complement and replace the current generations. Accessing information anywhere, anytime, with a seamless connection to a wide range of information and services, and receiving a large volume of information, data, pictures, video, and so on, are the keys features of 4G. Based on the developing trends of mobile communication, 4G will have broader bandwidth, higher data rate, and smoother and quicker handoff to provide seamless service across a multitude of wireless systems and networks. One of the major issues of seamless mobility is handoff management. It is a major challenge to design intelligent handoff management schemes for 4G-systems. In this paper we have presented the design of an adaptive multi-attribute vertical handoff decision algorithm based on genetic algorithm which is both cost effective and useful.

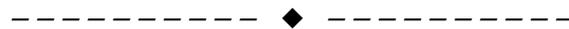

## 1. INTRODUCTION

Mobile wireless industry has started its technology creation, revolution and evolution since early1970s. In the past few decades, mobile wireless technologies have experienced 4 generations of technologies revolution and evolution, namely from 1G to 4G. The cellular concept was introduced in the 1G technology which made the large scale mobile wireless communication possible. Digital communication has replaced the analogy technology in the 2G which significantly improved the wireless communication quality. Data communication, in addition to the voice communication, has been the main focus in the 3G technologies and a converged network for both voice and data communication is emerging. With continued R&D, there are many killer application opportunities for the 4G as well as technological Challenges. In this paper we focus on one of such challenge that is vertical handoff management in 4G networks.

## 2. HANDOFF METRICS

Handoff metrics are the qualities that are measured to give an indication of whether or not a handoff is needed. As stated previously, in traditional handoffs only signal strength and channel availability are considered. In the envisioned **4G** system, the following new metrics have been proposed for use in conjunction with signal strength measurements:

**Service type.** Different types of services require various combinations of reliability, latency**,** and data rate.

**Monetary cost.** Cost is always a major consideration to users, as different networks may employ different billing strategies that may affect the user's choice of handoff.

**Network conditions.** Network-related parameters such as traffic, available bandwidth, network latency, and congestion (packet loss) may need to be considered for effective network usage. Use of network information in the choice to hand off can also be useful for load balancing across different networks, possibly relieving congestion in certain systems.

**System performance.** To guarantee the system performance, a variety of parameters can be employed in the handoff decision, such as the channel propagation characteristics, path **loss,** interchange interference, signal-to-noise ratio (SNR), and bit error rate (BER). In addition, battery power may be another crucial factor for certain users. When the battery level is low, the user may choose to switch to a network with lower power requirements, such as an ad hoc Bluetooth network.

### 2.1 Vertical handover in 4g networks

The integration of cellular networks and wireless local area networks is the development trend of the next generation mobile communication systems. Compared with homogeneous networks, the heterogeneous networks have the advantages of wider coverage, lower cost, higher rate and better quality of service (QoS). During the course of network integrations, an urgent and important issue that needs to be resolved is the mobility management or connection management [1] Mobility management contains two aspects: location management and handoff management [2]. Location management is used to track the positions of nodes no matter that they are communicating, and handoff management is used to keep

- *Bijay Kumar Paikaray is with CSE Dept, Centurion University of Technology and Management, Bhubaneswar Campus, Orissa, India.*

- *Er. Ajit Kumar Pasayat is with CSE Dept. Centurion University Of Technology and Management, Bhubaneswar Campus, Orissa, India*

communication uninterrupted when active nodes handoff from one cell to the other. The handoff decision algorithm belongs to handoff management is the research focus, which consists of horizon handoff decision [3, 4] that nodes handoff in homogeneous networks and vertical handoff decision [5 - 12] that nodes handoff in heterogeneous networks. In this paper, we pay attention to vertical handoff decision algorithm.

The algorithms concerned to the vertical handoff recorded in literatures can be classified to four main directions. The first approach is based on the strategies of using the received signal strength Indicator (RSSI) that may be combined with other parameters such as the left bandwidth [5 - 7]. The second approach utilizes artificial intelligence techniques such as neural network and fuzzy control combined several parameters such as RSS, network conditions and moving speed to make handoff decision [8, 9]. The third direction combines several metrics such as bandwidth, delay, access cost and power consumption in a cost function estimated for the available access networks, which is used for handoff decision [10, 11]. Three above approaches are only take the performance of nodes into consideration and they do not consider the system performance. The fourth approach solve the limitation of the above three methods, which obtains some optimal thresholds to maximize the system performance and there exist very few works related to it. A vertical handoff algorithm to optimize the quality of grade (GoS) of the system is proposed [12], in which an optimal velocity threshold is acquired to maximize the GoS. When the speed is lower than the threshold, the node will handoff to WLAN, otherwise, it stays in cell. There are many factors which affect handover. However, in this paper, we only take the RSS into consideration. This is because: 1) the ping-pong effects come from the characteristic of signal fading, so only take an effective method can reduce the ping-pong number; 2) Albeit the handoff decision is effected by many factors, the RSSI is a very important one and other factors are functions of it; 3) RSSI can easily be combined with other methods such as cost function, neural networks and fuzzy controlling to make more reasonable decision.

## 2.2 Basics of genetic algorithm
### 2.2.1 Genetic Algorithm (1) – Search space

Most often one is looking for the best solution in a specific subset of solutions.This subset is called the **search space** (or state space).Every point in the search space is a possible solution.Therefore every point has a **fitness** value, depending on the problem definition.GA's are used to search the search space for the best solution, e.g. a minimum.Difficulties are the local minima and the starting point of the search**.**

### 2.2.2 Genetic Algorithm (2) – Basic algorithm
Starting with a subset of *n* randomly chosen solutions from the search space (i.e. chromosomes). This is the population. This population is used to produce a next generation of individuals by reproduction.Individuals with a higher fitness have more chance to reproduce (i.e. natural selection)

### 2.2.3 Genetic Algorithm (3) – Basic algorithm
Basic algorithm
0  START   : Create random population of n chromosomes
1  FITNESS : Evaluate fitness f(x) of each chromosome in the population
2  NEW POPULATION
0 SELECTION    : Based on f(x)
1 RECOMBINATION : Cross-over chromosomes
2 MUTATION    : Mutate chromosomes
3 ACCEPTATION   : Reject or accept new one
3 REPLACE : Replace old with new population: the new generation
4  TEST  : Test problem criterium
5 LOOP : Continue step 1–4 until criterium is satisfied.

### 2.2.4 Simple Genetic Algorithm
```
{
initialize population;
evaluate populatSimple Genetic Algorithmion;
while TerminationCriteriaNotSatisfied
{
select parents for reproduction;
perform recombination and mutation;
evaluate population;
}
}
```

## 2.3. Mutation operation.
Select 1 parent probabilistically based on fitness. Pick point from 1 to NUMBER-OF-POINTS. Delete subtree at the picked point .Grow new subtree at the mutation point in same way as generated trees for initial random population (generation 0).The result is a syntactically valid executable program. Put the offspring into the next generation of the population.

## 2.4 Crossover operation
Select 2 parents probabilistically based on fitness .Randomly pick a number from 1 to NUMBER-OF-POINTS for 1st parent. Independently randomly pick a number for 2nd parent. The result is a syntactically valid executable program. Put the offspring into the next generation of the population. Identify the sub trees rooted at the two picked points.

## 2.5 Reproduction operation
Select parent probabilistically based on fitness .Copy it (unchanged) into the next generation of the population

## 3. VERTICAL HANDOFF DECISION FOR NEXT GENERATION WIRELESS NETWORK

The next generation wireless networks (NGWNs), called beyond third generation (B3G) or fourth generation (4G), will include multiple complementary mobile and wireless technologies, all of which will coexist in a heterogeneous wireless access environment and use a common IP core to offer a diverse range of high data rate multimedia services to end users. Because no single access technology can provide ubiquitous coverage and continuously high quality of service (QoS), multimode mobile terminals will have to roam among the various access technologies to maintain network connectivity and user satisfaction. These heterogeneous wireless access networks typically differ in terms of signal strength, coverage, data rate, latency, and loss rate. Therefore, each of them is practically designed to support a different set of specific services and devices. The limitations of these complementary wireless access networks can be overcome through the integration of the different technologies into a single unified platform (that is, an NGWN) that will empower mobile users to be connected to the NGWN using the best available access network that suits their needs. For example, third-generation (3G) cellular wireless wide area networks (WWANs) such as Universal Mobile Telecommunications System (UMTS) are designed primarily for mobile voice and data users whilst IEEE 802.16 or WiMAX (World-wide Interoperability for Microwave Access) systems are optimized to provide high-rate broadband wireless connectivity in wireless metropolitan area network (WMAN) environments for services and applications that require quality of service (QoS) guarantees. In addition, a mobile WiMAX overlay to a 3G wireless system can provide mobile operators with low cost additional capacity in spectrum and infrastructure limited regions, new real-time high speed data services, a proven and available path to an all-IP future and a seamless end user experience across services. The integration and internetworking of the heterogeneous wireless access networks in the NGWN require the design of intelligent vertical handoff decision algorithms (VHDAs) to enable mobile users to seamlessly switch network access and experience uninterrupted service continuity anywhere and anytime. Vertical handoff decision involves a tradeoff among many handoff metrics including QoS requirements (such as network conditions and system performance), mobile terminal conditions, power requirements, application types, user preferences, and a price model. Using these metrics involves the optimization of key parameters (attributes), including signal strength, network coverage area, data rate, reliability, security, battery power, network latency, mobile velocity, and service cost. These parameters may be of different levels of importance to the vertical handoff decision. Some categories of vertical handoff decision algorithm are proposed in the current research literature. The first category is based on the traditional strategy of using the received signal strength (RSS) combined with other parameters. In [1], Ylianttila et al. show that the optimal value for the dwelling timer is dependent on the difference between the available data rates in both networks. Another category uses a cost function as a measurement of the benefit obtained by handing off to a particular access network. In [2], it has been proposed a policy-enabled handoff across a heterogeneous network environment using a cost function defined by different parameters such as available bandwidth, power consumption, and service cost. The cost function is estimated for the available access networks and then used in the handoff decision of the mobile terminal (MT). Using a similar approach as in [2], a cost function-based vertical handoff decision algorithm for multi-services handoff was presented in [3]. The available network with the lowest cost function value becomes the handoff target. However, only the available bandwidth and the RSS of the available networks were considered in the handoff decision performance comparisons. The third category of handoff decision algorithm uses multiple criteria (attributes and/or objectives) for handover decision. An integrated network selection algorithm using two multiple attribute decision making (MADM) methods, Analytic Hierarchy Process (AHP) and Grey Relational Analysis (GRA), is presented in [4] with a number of parameters. In [5], Nasser et al. propose a vertical handoff decision function that provides handoff decision when roaming across heterogeneous wireless networks. However, computational intelligence techniques were not used. The fourth category of vertical handoff decision algorithm uses computational intelligence techniques. In [6], Pahlavan et al. present a neural networks-based approach to detect signal decay and making handoff decision. In [7], Chan et al. propose a mobility management in a packet-oriented multi segment using Mobile IP and fuzzy logic concepts. Fuzzy logic is applied to the handover initiation phase, and fuzzy logic and multiple objective decision making concepts are applied during the decision phase to select an optimum network. However, the handover management is for vertical handoff between different wide area networks. In this paper, we present the design of a fuzzy logic based vertical handoff initiation scheme involving some key parameters, and the solution of the wireless network selection problem using a fuzzy multiple attribute decision making (FMADM) access network selection function that is optimized with a Genetic Algorithm (GA) to select an optimum wireless access network.

### 3.1 Overview of the VHDA

Vertical handoff decision in an NGWN environment is more complex and involves a tradeoff among many handoff metrics including QoS requirements (such as network conditions and system performance), mobile terminal conditions, power requirements, application types, user preferences, and a price model. .A vertical handoff decision must solve the following problem: given a mobile user equipped with a contemporary multi-interfaced mobile device (with radio interfaces including UMTS, WLAN, WiMAX, and Digital Video Broadcasting-Handheld) connected to an access network, determine whether a vertical handoff should be initiated and dynamically select the optimum network connection from the available access network technologies to continue with an existing service. Consequently, our proposed VHDA consists of two parts [8]: a Handoff Initiation Algorithm and an Access Network Selection Algorithm.

The vertical handoff decision function is triggered when any of the following events occur: (a) when the availability of a new attachment point or the unavailability of an old one is detected, (b) when the user changes his/her profile, and thus altering the weights associated with the network selection attributes, and (c) when there is severe signal degradation or complete signal loss of the current radio link. Then the two part algorithm is executed for the purpose of finding the optimum access network for the possible handoff of the already running services to the optimum target network.

The Handoff Initiation Algorithm uses a fuzzy logic inference system (FIS) in figure-2 to process multiple vertical handoff initiation parameters (criteria). We use a Mamdani FIS that is composed of the functional blocks [9]: a *fuzzifier*, a *fuzzy rule base*, a *database*, a *fuzzy inference engine*, and a *defuzzifier*.

Since the inputs and outputs of the FIS are crisp in nature, the fuzzifier and defuzzifier are needed to transform them to and from fuzzy representation.

The Access Network Selection (ANS) algorithm involves decision making in a fuzzy environment. It can be solved using fuzzy multiple attribute decision making (FMADM) which deals with the problem of choosing an alternative from a set of alternatives based on the classification of their imprecise attributes. It applies a multiple attribute defined access network selection function (ANSF) to select the best access network that is optimized to the user's location, device conditions, service and application requirements, cost of service and throughput. This paper proposes to use a GA to optimize the ANSF with the goal of selecting the optimal access network.

The block diagram shown in Figure-1 describes the vertical handoff decision algorithm.

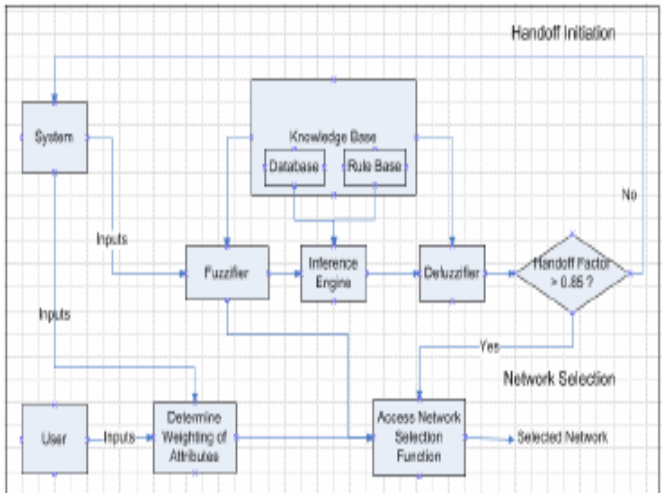

**Figure-1. Block diagram for Vertical Handoff Decision Algorithm.**

### 3.2. Handoff Initiation Algorithm

Computing and choosing the correct time to initiate vertical handoff reduces subsequent handoffs, improves QoS, and limits the data signaling and rerouting that is inherent in the handoff process. To process vertical handoff-related parameters, we use fuzzy logic, which uses approximate modes of reasoning to tolerate vague and imprecise data that are affected by errors in precision and accuracy from measurements by the system. An FIS expresses mapping rules in terms of linguistic language. A Mamdani FIS can be used for computing accurately the handoff factor which determines whether a handoff initiation is necessary between an UMTS and WiMAX. We consider two handoff scenarios: handoff from UMTS to WiMAX, and handoff from WiMAX to UMTS.

### 3.2.1. Handoff from UMTS to WiMAX

Suppose that an MT that is connected to a UMTS network detects a new WiMAX network. An FIS calculates the handoff factor which determines whether the MT should handoff to the WiMAX. The input parameters are the RSS indication (RSSI), data rate, network coverage area, and network latency of the target WiMAX network. The crisp values of the input parameters are fed into a fuzzifier in a Mamdani FIS, which transforms them into fuzzy sets by determining the degree to which they belong to each of the appropriate fuzzy sets via membership functions (MFs). Triangular, trapezoidal, and Gaussian functions are typically used as MFs. Next, the fuzzy sets are fed into a fuzzy inference engine where a set of fuzzy IF-THEN rules is applied to obtain fuzzy decision sets. The output fuzzy decision sets are aggregated into a single fuzzy set and passed to the defuzzifier to be converted into a precise quantity, the handoff factor, which determines whether a handoff is necessary.

Each of the input parameters is assigned to one of three fuzzy sets; for example, the fuzzy set values for the RSSI consist of the linguistic terms: Strong, Medium, and Weak. These sets are mapped to corresponding Gaussian MFs. The universe of discourse for the fuzzy variable RSSI is defined from -78 dBm to -66 dBm. The universe of discourse for the variable Data Rate is defined from 0 Mbps to 60 Mbps, the universe of discourse for the variable Network Coverage is defined from 0 m to 50 Km, and the universe of discourse for the variable Network Latency is defined from 0 to 200 ms. The fuzzy set values for the output decision variable Handoff Factor are Higher, High, Medium, Low, and Lower. The universe of discourse for the variable Handoff Factor is defined from 0 to 1, with the maximum membership of the sets "Lower" and "Higher" at 0 and 1, respectively. Since there are four fuzzy input variables and three fuzzy sets for each fuzzy variable, the maximum possible number of rules in our rule base is = 81. The fuzzy rule base contains IF-THEN rules such as:

IF RSSI is weak, and data rate is low, and network coverage area is bad, and network latency is high, THEN handoff factor is lower.

IF RSSI is strong, and data rate is high, and network coverage area is good, and network latency is low, THEN handoff factor is higher. The crisp handoff factor computed after defuzzification is used to determine when a handoff is required as follows:

if *handoff factor* > 0.65, then initiate handoff; otherwise do nothing.

### 3.2.2 Handoff from WiMAX to UMTS

The parameters that we are using in this directional handoff include the RSSI, data rate, network coverage area, and network latency of the current WiMAX network. The design of the fuzzy inference system for this handoff scenario is similar to the design of the fuzzy inference system for the UMTS-to-WiMAX handoff. However, the fuzzy rule base contains IF-THEN rules which are the direct opposite of those in the UMTS-to-WiMAX handoff.

### 3.2.3 Network Selection Algorithm

A suitable access network has to be selected once the handoff initiation algorithm indicates the need to handoff from the current access network to a target network. The network selection decision process is formulated as a FMADM problem that deals with the evaluation of a set of alternative access networks using a multiple attribute access network selection function (ANSF) defined on a set of attributes. Fuzzy logic is used to deal with imprecise data that the decision attributes might contain, and to combine and evaluate multiple attributes simultaneously. The ANSF is an objective function that measures the efficiency in utilizing radio resources and the improvement in QoS to mobile users gained by handing off to a particular network. It is defined for all alternative target access networks that cover the service area of a user. The network that provides the highest ANSF value is selected as the best network to handoff from the current access network according to the mobile terminal conditions, network conditions, service and application requirements, cost of service, and user preferences.

The ANSF is triggered when any of the following events occur: (a) a new service request is made; (b) a user changes his/her preferences; (c) the MT detects the availability of a new network; (d) there is severe signal degradation or complete signal loss of the current radio link. Parameters (attributes) used for the ANSF include the signal strength (*S*), network coverage area (*A*), data rate (*D*), service cost (*C*), reliability (*R*), security (*E*), battery power (*P*), mobile terminal velocity (*V*), and network latency (*L*). The main objective of the network selection algorithm is to determine and select an optimum cellular/wireless access network for a particular high-quality service that can satisfy the following goals:

**Good signal strength**: Signal strength is used to indicate the availability of a network, and an available network can be detected if its signal strength is good.

**Good network coverage**: Frequent handoffs incur delay and loss of packets. A network that provides a large coverage area enables mobile users to avoid frequent handoffs as they roam about.

**Optimum data rate**: A network that can transfer signals at a high rate is preferred since a maximum data rate reduces service-delivery time for non-real time services and enhances QoS for adaptive real time services.

**Low service cost**: The cost of services offered is a major consideration to users and may affect the user's choice of access network and consequently handoff decision. A user may prefer to be connected through the cheapest available access network in order to reduce service cost incurred.

**High reliability**: A reliable network is not error prone and so can be trusted to deliver a high level of performance.

**Strong security**: As strong security enhances information integrity, a network with high encryption is preferred when the information exchanged is confidential.

**Good MT velocity**: Handing off to an embedded network in an overlaid architecture of heterogeneous networks is discouraged when traveling at a high speed since a handoff back to the original network will occur very shortly afterward when the mobile terminal leaves the smaller embedded network. High mobile users are connected to the upper layers and benefit from a greater coverage area.

**Low battery power requirements**: Power consumption should be minimized since mobile devices have limited power capabilities. When the battery level decreases, handing off to a network with lower power requirements would be a better decision.

**Low network latency**: High network latency degrades applications and the transfer of information. A handoff algorithm should be fast so that the mobile device does not experience service degradation or interruption.

The optimum wireless access network must satisfy:
$$\text{Maximize } f(u)u$$

Where $f_i(u)$ the objective function is evaluated for the network I and u is the vector of input parameters. The function $f_i$ can be expressed as:

$$f_i(u)=f(S,A,D,1/C_i,R_i,E_i,V_i,1/P_i,1/L_i)$$
$$=\sum_{i=1}^{6} W_x \cdot N_i(X_i)+\sum_{i=1}^{3} w_y \cdot N_f(1/Y_i)$$

Where $N_f(X)$ is the normalized function of the parameter X and $W_x$ is the weight which indicates the importance of the parameter X, with $X_i= S_i,A_i,D_i, R_i,E_i,V_i$ and $Y_i=C_i,P_i,L_i$``

$$f_i(x)=\sum_{i=1}^{6} W_x \cdot (X_i/X_{max}))+\sum_{i=1}^{3} w_y \cdot Y/Y_{max})$$

A suitable normalized function of the parameter *X* is the fuzzy membership function μ(x). In order to develop this function, data from the system are fed into a fuzzifier to be converted into fuzzy sets. The values of the parameters are normalized between 0 and 1. Then a single membership function is defined such that $\mu_{Cj}(0) = 0$ and $\mu_{Cj}(1) = 1$ if the goal is to select a network with a high parameter *X* value; and such that $\mu_{Cj}(0) = 1$ and $\mu_{Cj}(1) = 0$ if the goal is to select a network with a low parameter *X* value.

In general, the fitness value for the network *i* is thus given by

$$f_i(x)= \sum_{i=1}^{n} W_j \cdot \mu_{Cj}(A_i),$$

Where x is the vector of membership function values.

The optimum wireless network is given by the optimization problem:

$$\text{Max } f_i(x) = \max \{ \sum_{i=1}^{n} W_j \cdot \mu_{Cj}(A_i),\}$$

Such that
$$0<W_j<1, \text{ and } \sum_{j=1}^{n} W_j =1$$

And
$$\{ \mu_{Cj}(A_i),\}< \mu_{Cj}(A_i)<\{ \mu_{Cj}(A_i),\}$$

The last inequality constraint arises from the fact that every attribute is bounded either by an acceptable minimum threshold value and the possible maximum value, or by the possible minimum value and an acceptable maximum threshold value.

**Determination of Attribute Weights**: Data from the system are fed into a fuzzifier to be converted into fuzzy sets. Suppose that $A = \{A1, A2, \ldots, Am\}$ is a set of *m* alternatives and $C = \{C1, C2, \ldots, Cn\}$ is a set of *n* handoff decision criteria (attributes) that can be expressed as fuzzy sets in the space of alternatives. The criteria are rated on a scale of 0 to 1. The degree of membership of alternative *Aj* in the criterion *Ci*, denoted $\mu_{Cj}(A_i)$, is the degree to which alternative *Aj* satisfies this criterion. A decision maker (such as a mobile user) judges the criteria in pair wise comparisons [10], and assigns the values *aij* = 1/*aji* using the judgment scale proposed by Saaty: 1– equally important; 3 – weakly more important; 5 – strongly more important; 7 – demonstrably more important; 9 – absolutely more important. The values in between {2, 4, 6, 8} represent compromise judgments. An *n* x *n* matrix Bi constructed so that:

(1) $b_{ii} = 1$; (2) $b_{ij} = a_{ij}, i \neq j$; (3) $b_{ji} = 1/b_{ij}$.

Using this matrix, the unit eigenvector, V, corresponding to the maximum eigenvalue, $\lambda max$, of B is then determined by solving the equation:

$$B \cdot V = \lambda max \cdot V \quad (7)$$

Finding the unit eigenvector, V, corresponding to the maximum eigenvalue of B produces the cardinal ratio scale of the compared attributes. The eigenvectors are then normalized to ensure consistency. In other words, the values of V are scaled for use as factors in weighting the membership values of each attribute by a scalar division of V by the sum of values of V to obtain a weighting matrix W. The MT calculates the handoff initiation factor in the handoff initiation algorithm when the MT detects a new network or the user changes his/her preferences or the current radio link is about to drop. If the handoff initiation algorithm indicates the need for a handoff of the already running services from the current network to a target network, the mobile terminal then calculates the ANSF $f_i$ for the current network and target networks. Vertical handoff takes place if the target network receives a higher $f_i$.

### 3.3 GA Optimization of the ANSF

In this section we explore the use of GAs for solving the optimization problem of maximizing the ANSF in equation (4). The GA is a search method for solving both constrained and unconstrained optimization problems that is based on the principles of natural selection and genetics [11]. GAs operates on encoded representations of the solutions (also called chromosomes) rather than the solutions themselves. Each solution is associated with a fitness measure that reflects how good it is, compared with other solutions in the population. The measure could be an objective function that is a mathematical model or a computer simulation. In the following, we assume a function minimization problem. Hence, a good solution is one that has low relative fitness. Once a problem is encoded in a chromosomal manner and a fitness measure has been chosen, we can use a GA to evolve solutions to the problem by the following steps:

**Step 1:** *Initialization*. The algorithm begins by creating a random initial population of candidate solutions.

**Step 2:** The algorithm then creates a sequence of new populations. At each step, the algorithm uses the individuals in the current generation to create the next population by iteratively performing the following steps:

a) *Evaluation.* The objective function values of the candidate solutions in the current population are evaluated.

b) *Fitness Assignment.* The algorithm uses the objective function values to determine the fitness values of the candidate solutions in the current population.

c) *Selection.* The algorithm selects members, called parents, based on their fitness. The main idea of selection is to prefer better solutions to worse ones.

d) *Elitism.* Some of the individuals in the current population that have the best fitness values are chosen as *elite* individuals and are passed to the next population as children.

e) *Crossover (Recombination).* Crossover combines the vector entries or genes of two parents to form potentially better solutions (offspring) for the next generation. The crossover is controlled by the crossover probability *pc* which is typically in the range [0.7 – 0.95].

f) *Mutation.* Mutation applies random changes to one or more genes of an individual parent to form children. It is performed with a low probability *pm* in the range [0.01 – 0.2].

g) *Reproduction.* Reproduce the children created by selection, crossover, and mutation to form the next generation.

**Step 3:** The algorithm stops when a stopping criteria is met. To tackle the optimization problem of maximizing the ANSF in equation (4) under the weights $w_j$ by using a GA, we assume a function minimization problem. Hence, a good solution is one that has low relative fitness. Since our GA algorithm performs minimization of an objective function $f(x)$, maximization of the objective function in equation (4) is achieved by supplying the routine with minus $f_i(x)$ because the point at which the minimum of $-f_i(x)$ occurs is the same as the point at which the maximum of $f_i(x)$ occurs. Therefore, we define the equivalent minimization problem: In order to achieve good results in using a GA, one may have to experiment with different GA operators (selection, crossover, and mutation operators) and different GA parameters (population size, crossover probability, and mutation probability).

## 4. PERFORMANCE EVALUATION OF NETWORK SELECTION

Three cellular networks (GPRS_1, UMTS_1, and UMTS_2) cover the entire metropolitan area. Two IEEE 802.16e WMAN networks (WiMAX_1 and WiMAX_2) partly overlay the inner city Figure-3. Suppose a person is using UMTS_1 currently. He is downloading some files in its current network. Suppose he enters to another network suppose WiMAX_1.Then these following steps will evaluate whether handoff occurs or not in Figure-4.

*Evaluation:* We first check to see whether a handoff should be initiated by calculating the handoff initiation factor. Suppose that the MT records the data values of RSSI (dBm), Data Rate (Mbps), Network Coverage Area (Km), and Network Latency (ms) as {-67.3, 48.8, 47.9, 56.5} and {- 67.01, 48.6, 47.6, 55.8} for WiMAX_1 and WiMAX_2 respectively. These set of values are fed into the FIS and we obtain the Handoff Factor values 0.75 and 0.76, thus indicating the need to hand off to any of the WMANs for the requested service. The second stage of the VHDA is to compute the ANSF for all the available networks. The mobile terminal proceeds to gather data on all required parameters. The matrix B and weighting matrix W are indicated below: (9) the attribute weights and the membership values (lower bound, upper bound) obtained from the characteristics of the three available networks for the attributes are summarized in Table 2.

| Criteria | | wj | Membership Values(lb, ub) | | |
|---|---|---|---|---|---|
| | | | UMTS_2 | WiMAX_1 | WiMAX_2 |
| RSSI | $c_1$ | 0.0181 | 0.5,0.9 | 0.5,0.9 | 0.5,0.9 |
| Data Rate | $c_2$ | 0.4772 | 0.05,0.1 | 0.2,0.9 | 0.2,0.9 |
| Network Coverage | $c_3$ | 0.0242 | 0.1,0.45 | 0.2,0.95 | 0.2,0.92 |
| Network Latency | $c_4$ | 0.1198 | 0.4,0.6 | 0.5,0.9 | 0.5,0.9 |
| Reliability | $c_5$ | 0.0685 | 0.7,0.9 | 0.7,0.9 | 0.7,0.9 |
| Security | $c_6$ | 0.0181 | 0.8,0.9 | 0.8,0.9 | 0.8,0.9 |
| Power Requirement | $c_7$ | 0.00181 | 0.7,0.8 | 0.6,0.75 | 0.6,0.75 |
| Mobile Velocity | $c_8$ | 0.0385 | 0.01,0.9 | 0.01,0.5 | 0.01,0.5 |
| Service Cost | $c_9$ | 0.2202 | 0.5,0.6 | 0.6,0.83 | 0.6,0.85 |

Table 2: Different membership values for different networks before optimization.

| Criteria | | wj | Membership Values(lb, ub) | | |
|---|---|---|---|---|---|
| | | | UMTS_2 | WiMAX_1 | WiMAX_2 |
| RSSI | $c_1$ | 0.0181 | 0.8125 | 0.8945 | 0.9110 |
| Data Rate | $c_2$ | 0.4772 | 0.0994 | 0.9000 | 0.9000 |
| Network Coverage | $c_3$ | 0.0242 | 0.2027 | 0.8039 | 0.8879 |
| Network Latency | $c_4$ | 0.1198 | 0.5949 | 0.7839 | 0.8865 |
| Reliability | $c_5$ | 0.0685 | 0.9000 | 0.9000 | 0.8898 |
| Security | $c_6$ | 0.0181 | 0.8985 | 0.8938 | 0.8993 |
| Power Requirement | $c_7$ | 0.00181 | 0.7998 | 0.7484 | 0.6552 |
| Mobile Velocity | $c_8$ | 0.0385 | 0.8972 | 0.5000 | 0.5000 |
| Service Cost | $c_9$ | 0.2202 | 0.5982 | 0.8300 | 0.8500 |

Table 3: Different membership values for different networks after optimization.

We performed the GA optimization experiments by using the stochastic universal selection rule that lays out a line in which each parent corresponds to a section of the line of length proportional to its scaled value, and settled on the options: population size *ps* = 20, elite count = 2, single-point crossover with *pc* = 0.8, and mutation probability *pm* = 0.01. The solutions (a list of the optimum ANSF values and optimum membership function values) obtained using the GA (MATLAB GA toolbox) are summarized in Table 3. The best fitness value and the best individual for the WiMAX_2 network are plotted in Figure-5 And the MATLAB Optimization toolbox.

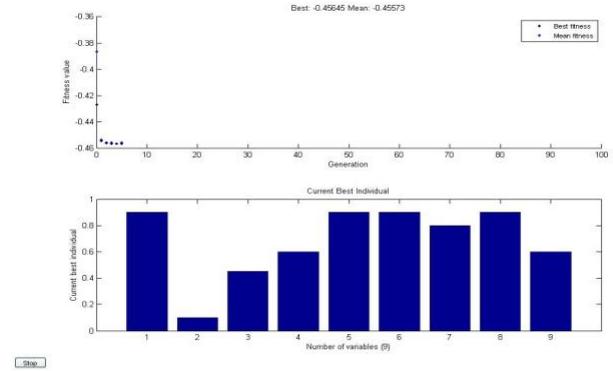

Fig 3: Performance evaluation for UMTS_2 network Using GA

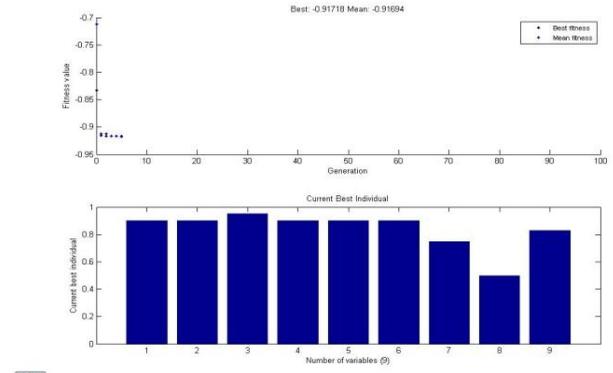

Fig 4: Performance evaluation for WiMAX_1 network Using GA

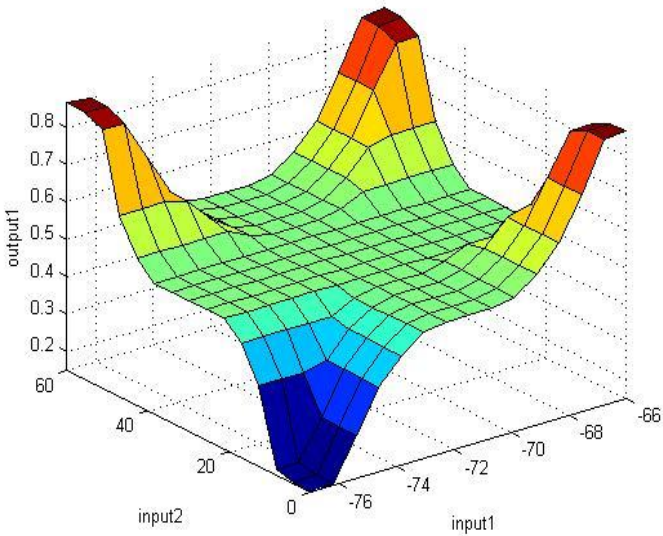

Fig 2: Representation of Fuzzy Inference System for Network

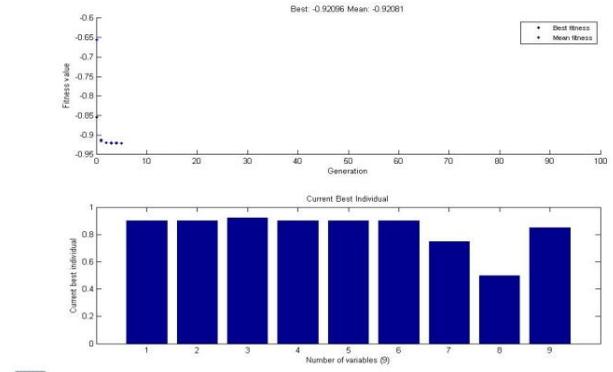

Fig 5: Performance evaluation for WiMAX_2 network Using GA.

## 5. CONCLUSION:

The convergence of wireless access technologies towards 4G wireless networks should overcome several challenges before practical implementation. One of the major challenges is user mobility handling between different accesses technologies in order to keep the user connected to the best available network.

A main challenge for seamless mobility in the evolving multi-access next generation wireless network is the design of intelligent handoff management schemes. In this paper we have presented the design of an adaptive multi-attribute vertical handoff decision algorithm based on genetic algorithms. We have compared the performance of the algorithm in different networks. The simulation result shows that it is both cost-effective and useful.

We demonstrated the use of fuzzy logic concepts to combine multiple attributes from the network to obtain useful handoff initiation schemes and used a GA to optimize the selection of suitable access networks with a fuzzy multiple attribute defined access network selection function.

## AUTHORS

**Mr. Bijay Paikaray** received his MCA from Sambalpur University in June 2010.Presently working as a Teach Asst. in the Dept. of CSE in Centurion University of Technology and Management, Bhubaneswar, India. His area of interest includes Wireless and Mobile Networks, Wireless Sensor networks, Computer Organization, Network Security.